\documentclass[aps,pre,twocolumn,amsmath,amssymb,superscriptaddress,floatfix,showpacs]{revtex4}

\usepackage{graphicx}
\usepackage{bm}
\usepackage[usenames]{color}
\bibstyle{apsrev.bib}

\newcommand{\be}{\begin{equation}}
\newcommand{\ee}{\end{equation}}
\newcommand{\beqn}{\begin{eqnarray}}
\newcommand{\eeqn}{\end{eqnarray}}

\begin{document}

\title{Long-range random transverse-field Ising model in three dimensions}
\author{Istv\'an A. Kov\'acs}
\email{kovacs.istvan@wigner.mta.hu}
\affiliation{Wigner Research Centre for Physics, Institute for Solid State
Physics and Optics, H-1525 Budapest, P.O.Box 49, Hungary}
\affiliation{Institute of Theoretical Physics, Szeged University, H-6720 Szeged,
Hungary}
\affiliation{Center for Complex Networks Research and Department of Physics,
Northeastern University, 177 Huntington Avenue,
Boston, MA 02115, USA}
\author{R\'obert Juh\'asz}
\email{juhasz.robert@wigner.mta.hu}
\affiliation{Wigner Research Centre for Physics, Institute for Solid State
Physics and Optics, H-1525 Budapest, P.O.Box 49, Hungary}
\author{Ferenc Igl\'oi}
\email{igloi.ferenc@wigner.mta.hu}
\affiliation{Wigner Research Centre for Physics, Institute for Solid State
Physics and Optics, H-1525 Budapest, P.O.Box 49, Hungary}
\affiliation{Institute of Theoretical Physics, Szeged University, H-6720 Szeged,
Hungary}
\date{\today}

\begin{abstract}
We consider the random transverse-field Ising model in $d=3$ dimensions with long-range ferromagnetic interactions which decay as a power $\alpha > d$ with the distance. Using a variant of the strong disorder renormalization group method we study numerically the phase-transition point from the paramagnetic side. The distribution of the (sample dependent) pseudo-critical points is found to scale with $1/\ln L$, $L$ being the linear size of the sample. Similarly, the critical magnetization scales with $(\ln L)^{\chi}/L^d$ and the excitation energy behaves as $L^{-\alpha}$. Using extreme-value statistics we argue that extrapolating from the ferromagnetic side the magnetization approaches a finite limiting value and thus the transition is of mixed-order.

\end{abstract}


\maketitle

\section{Introduction}
In nature there are magnetic materials in which ordering is due to long-range (LR) interactions which decay as a power $\alpha=d+\sigma$ with the distance. The best known examples are dipolar systems, such as the $\mathrm{LiHoF}_4$. Putting this compound into an appropriate external magnetic field we obtain an experimental realisation of a dipolar quantum ferromagnet\cite{LiHoF4}. Similar systems have been experimentally realised recently by ultracold atomic gases in optical lattices\cite{friedenauer,kim,islam,britton,islam1} and studied theoretically\cite{porras,deng,hauke,peter,nebendahl,wall,cannas,dutta,dalmonte,koffel,hauke1}.

Concerning the phase-transitional properties of LR systems it is known quite some time that the universality class depends on the decay
exponent, $\sigma$\cite{fisher_me,sak,bloete,picco,parisi}. For a sufficiently large value of $\sigma > \sigma_U$ the transition is the same as in the short-range (SR) model, for intermediate values, $\sigma_L >\sigma > \sigma_U$ the critical behaviour is non-universal and $\sigma$ dependent, while for $\sigma < \sigma_L$ we have mean-field critical behaviour. In low-dimensional systems LR forces could result in magnetic ordering and phase transitions, even if these are absent with SR interactions\cite{dyson}.

In the present paper we consider quantum magnets with LR interactions in the presence of quenched disorder. Such type of a system is realised
by the compound $\mathrm{LiHo}_x\mathrm{Y}_{1-x}\mathrm{F}_4$, in which a fraction of $(1-x)$ of the magnetic $\mathrm{Ho}$ atoms is replaced by non-magnetic $\mathrm{Y}$ atoms\cite{experiment,LiHoF4}. A related, but somewhat simplified\cite{rfield} quantum model which describes the low-energy properties of this system is the random transverse-field Ising model with LR interactions given by the Hamiltonian:
\be
{\cal H} =
-\sum_{i\neq j} \frac{b_{ij}}{r_{ij}^{\alpha}} \sigma_i^x \sigma_{j}^x-\sum_{i} h_i \sigma_i^z\;.
\label{eq:H}
\ee
Here, $\sigma_i^{x,z}$ are Pauli-matrices, $r_{ij}$ denotes the distance between site $i$ and $j$,  while the parameters $b_{ij}$ and transverse fields $h_i$ are i.i.d. quenched random variables drawn from some distributions $p_0(b)$ and $g_0(h)$, respectively. In the following we restrict ourselves to ferromagnetic models, so that $b_{ij}>0$ and $h_i>0$.

The properties of phase-transitions in random quantum Ising magnets with SR interactions are known with some extent\cite{fisher,2d,2dki,ddRG}, mainly due to strong disorder renormalization group (SDRG)\cite{mdh,im} studies. These results are then checked by numerical investigations in one-\cite{young_rieger,igloi_rieger} and two-dimensions\cite{pich,matoz_fernandez}. The main conclusion is that the critical behaviour of SR random quantum Ising magnets at any finite dimension is governed by a so called infinite-disorder fixed point (IDFP)\cite{danielreview,im}, at which the dynamics is ultra-slow: the length-scale $\xi$ and the time-scale $\tau$ is related as:
$\ln \tau \sim \xi^{\psi}$, where the exponent $\psi$ is dimension dependent. This is in contrast with the scaling behaviour at a conventional random fixed-point: $\tau \sim \xi^z$, thus the dynamical exponent, $z$ is formally infinite at an IDFP.

The low-energy properties of the Hamiltonian in Eq.(\ref{eq:H}) in one-dimension have already been studied through variants of the SDRG\cite{jki}, see also studies of the related hierarchical Dyson model\cite{cecile} and that in Ref\cite{stretched}. In contrast to the SR case the critical behaviour is found to be controlled by a strong-disorder fixed point, the critical dynamical exponent being finite: $z_c\simeq\alpha$. 
Qualitatively similar observations are found from the preliminary numerical SDRG results on the two-dimensional system in Ref.\cite{jki1}.

In this paper we extend these investigations to the experimentally more realistic three-dimensional system. Here we apply a numerical version of the SDRG method, which is expected to present physically correct results in the critical point approaching from the paramagnetic side. We study in details the distribution of the sample dependent critical points, the scaling behaviour of the magnetization and that of the low-energy excitations. The SDRG results are then interpreted in the frame of extreme-value statistics (EVS)\cite{galambos}, which is then used to make conjectures about the scaling behaviour of the different quantities at the ferromagnetic side of the transition point. Most surprisingly the average magnetization is expected to approach a finite limiting value, thus the transition is of mixed order.

The rest of the paper is organised as follows. In Sec.\ref{SDRG} we recapitulate the basic features of the SDRG method and discuss its particular form for LR systems. In Sec.\ref{numSDRG} we present our results of the numerical SDRG analysis, which is then interpreted within the frame of EVS
in Sec.\ref{EVS}. Our results are discussed in Sec.\ref{discussion}.

\section{SDRG method for LR interactions}
\label{SDRG}

In the SDRG method the decimation procedure is performed locally in the energy-space of the Hamiltonian. At each step of the renormalization the largest local parameter of the Hamiltonian is eliminated and between the remaining degrees of freedom new, renormalized parameters are calculated perturbatively. After decimating the strongest coupling, $J_{ij}$, (in our case $J_{ij}=\frac{b_{ij}}{r_{ij}^{\alpha}}$) the two sites $i$ and $j$ form a cluster
of spins in an effective transverse field: $\tilde{h}=h_i h_j/J_{ij}$. On the contrary, if the strongest transverse-field, $h_i$ is decimated, then between sites, $j$ and $k$, -- which are originally nearest neighbors to $i$ -- a new coupling is created: $\tilde{J}_{jk}={\rm max}(J_{jk},J_{ji}J_{ki}/h_i)$. Here in the last step the so called maximum rule is applied, which is essential in the fast algorithm\cite{ddRG} we use in the numerical calculation. The use of the maximum rule is correct in the paramagnetic phase and asymptotically correct at an IDFP. If the fixed-point under consideration is just a strong-disorder one, then the maximum rule is a good approximation, which generally does not modify the qualitative behaviour of the model. We note, however, that the maximum rule is certainly not correct in the ferromagnetic phase in particular for LR models.

\begin{figure}[th]
\begin{center}
\includegraphics[width=3.4in,angle=0]{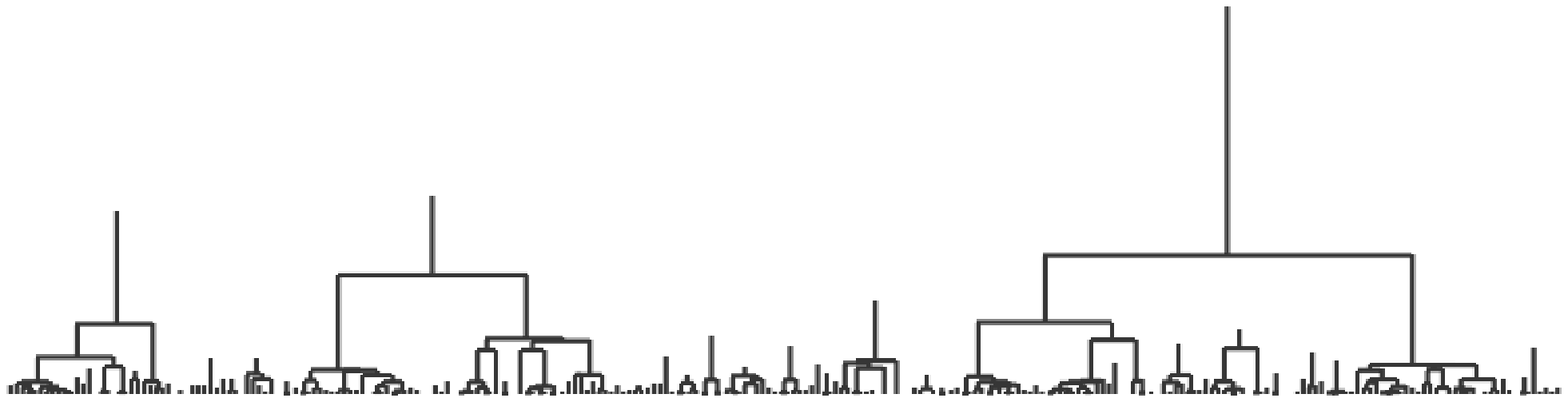}
\includegraphics[width=3.4in,angle=0]{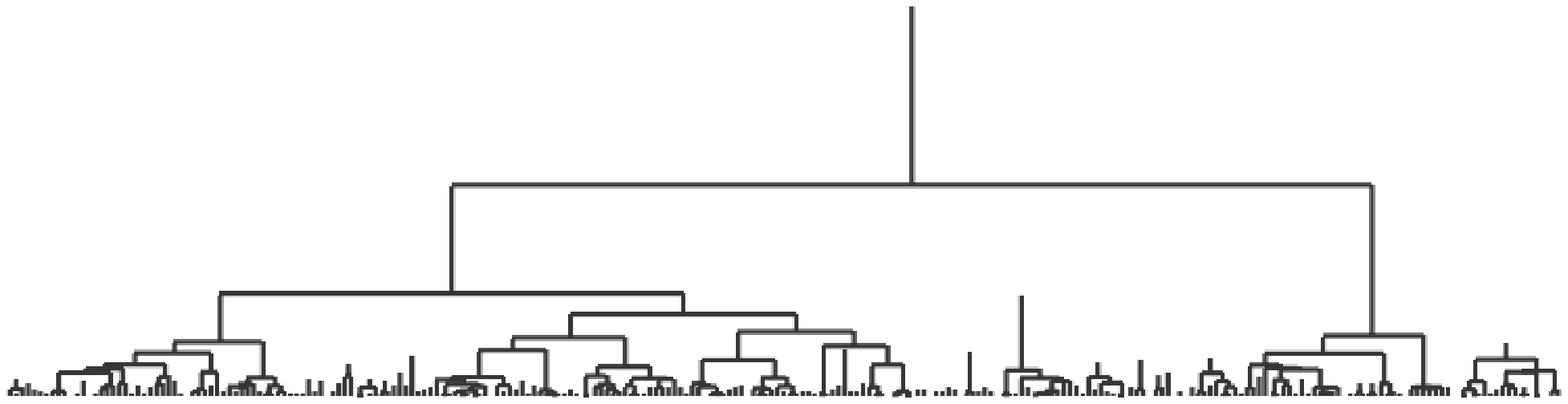}
\end{center}
\vskip -.5cm
\caption{
\label{fig_1} (Color online) Illustration of the spin clusters formed during the SDRG process at the critical point of the 1D short-range RTIM for two samples at $L=256$. The fate of a spin cluster is either to be decimated out (indicated by vertical spikes) or to be fused together with an other spin cluster (horizontal lines). Higher trees indicate clusters being present at later stages of the SDRG procedure, corresponding to the low-energy modes of the system. The magnetic moment is related to the size of the largest cluster, scaling as $\mu(L)\propto L^{d_f}$, where the fractal dimension is $d_f=\frac{\sqrt{5}+1}{4}\approx 0.809$\cite{fisher}.
}
\end{figure}

In the numerical application of the SDRG method we start with some initial disorder, in our case we have used box-distributions:
the parameters of the model were chosen uniformly from the intervals $b_{ij}\in(0,1]$ and $h_i\in(0,h]$, so that the control parameter is defined as $\theta=\ln(h)$. Now let us assume for a moment that our model is short-ranged, i.e. in Eq.(\ref{eq:H}) $\frac{b_{ij}}{r_{ij}^{\alpha}}$ is replaced with $b_{ij}$ and the first sum runs over nearest neighbours. At the critical point of the SR model $(\theta_c^{SR})$ the decimation procedure is asymptotically symmetric in 1D: couplings and transverse fields are decimated with the same fraction. The resulting cluster structure is illustrated in Fig.\ref{fig_1}. In higher dimensions the ratio of the frequency of coupling and transverse-field decimations has a finite limiting value, $r_{SR}={\cal O}(1)$. Now switching on the LR forces, the renormalization procedure starting from $\theta_c^{SR}$ will turn to be more and more asymmetric due to the appearance of new LR couplings: below some energy-scale almost always couplings will be decimated and the LR model renormalizes to the ferromagnetic fixed-point. Consequently the critical point of the LR model satisfies the relation $\theta_c > \theta_c^{SR}$.

\begin{figure}[th]
\begin{center}
\includegraphics[width=3.4in,angle=0]{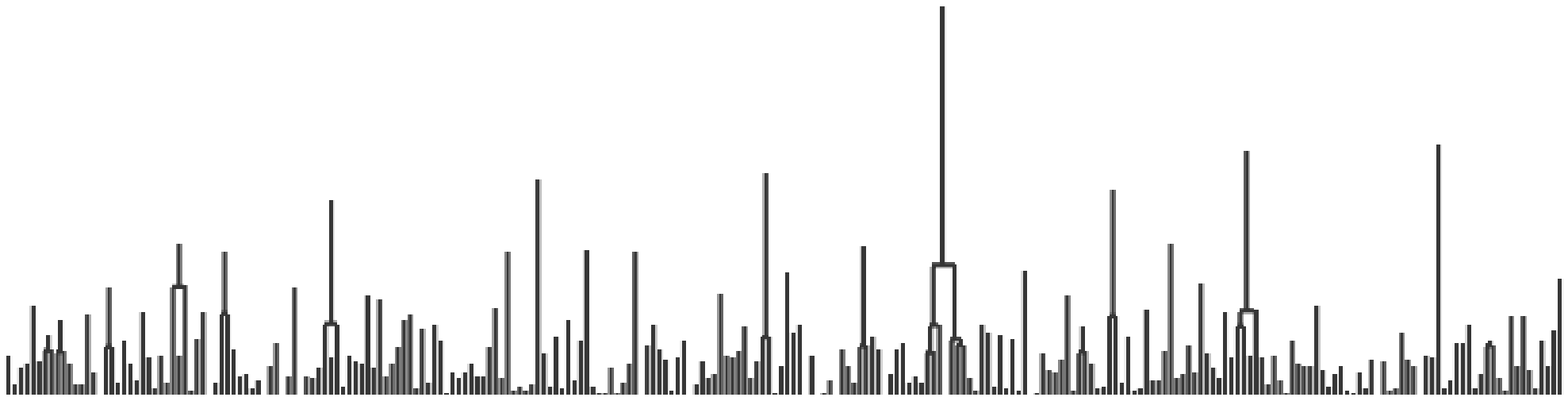}
\includegraphics[width=3.4in,angle=0]{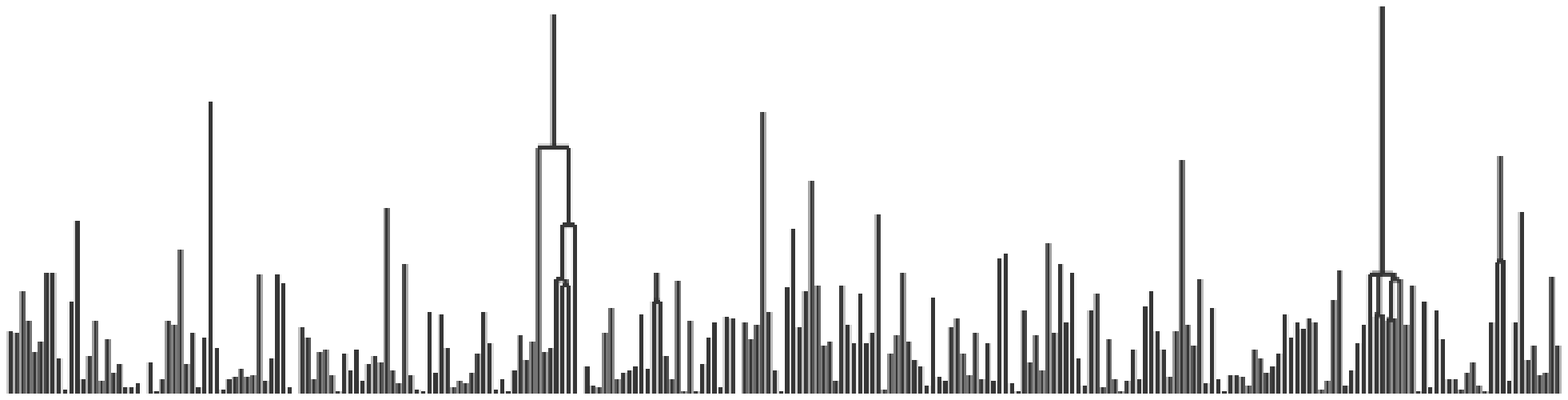}
\end{center}
\vskip -.5cm
\caption{
\label{fig_2} (Color online) The same dendrogram illustration as in Fig. \ref{fig_1} for the critical 1D long-range RTIM at $L=256$. As opposed to the short-range model, mostly transverse fields are decimated resulting in spikes and smaller spin-clusters, following a scaling of the form $\mu(L)\propto \ln^2 L$.
}
\end{figure}

\begin{figure}[h!]
\begin{center}
\vskip -2.5cm
\includegraphics[width=3.5in,angle=0]{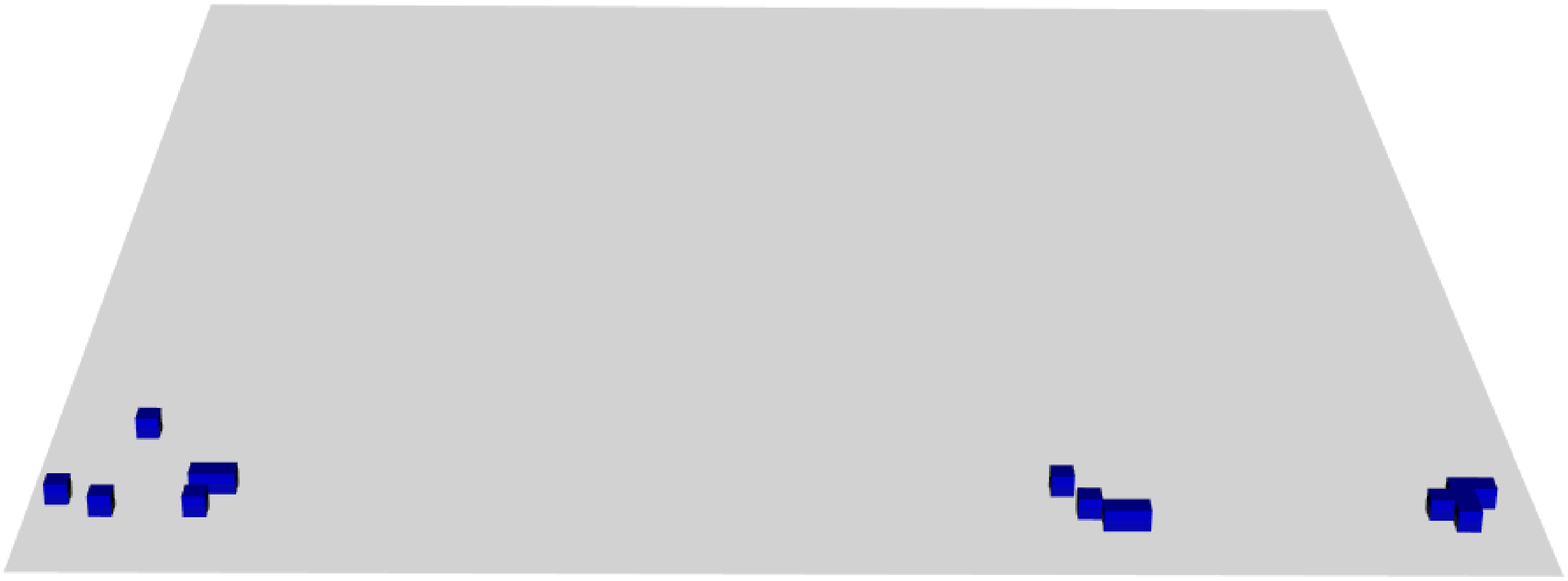}
\vskip -2.5cm
\includegraphics[width=3.2in,angle=0]{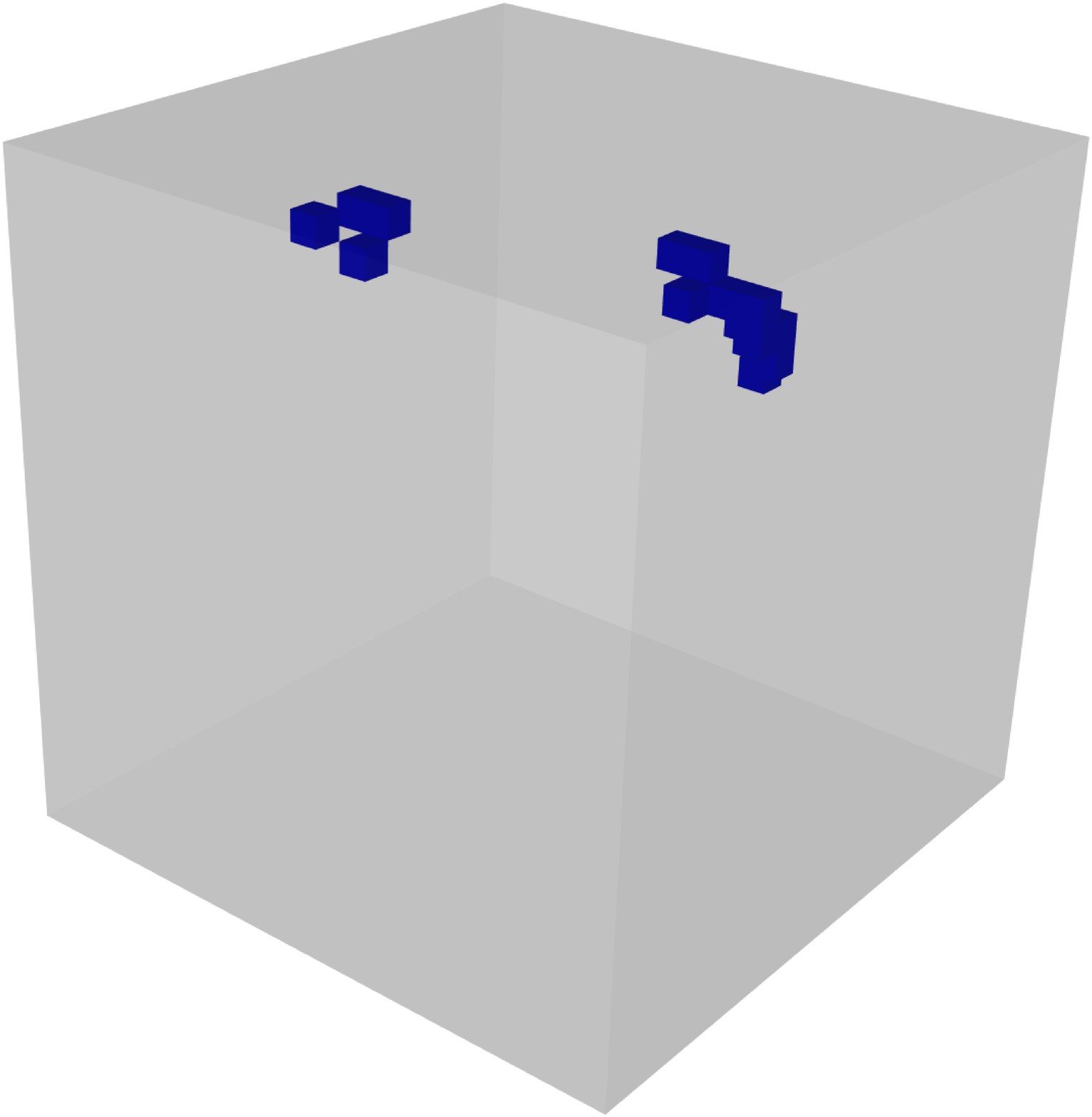}
\end{center}
\vskip -.5cm
\caption{
\label{fig_3} (Color online) The large scale spin clusters of the critical long-range RTIM appear to be sparse and they can be embedded in a quasi-one-dimensional object, as illustrated in 2D (L=64) and 3D (L=24). The size of these clusters provides the magnetic moment following a similar scaling as in 1D as shown in Eq. (\ref{mu_L}). 
}
\end{figure}

Now let us follow the renormalization procedure of the LR model starting from its own critical point $\theta_c$, when three different regimes can be identified. At the \textit{initial period} dominantly nearest-neighbour couplings are involved and the renormalization proceeds basically as in the SR model. Since at $\theta_c$ the SR model is in the paramagnetic phase in the initial period almost exclusively transverse fields are decimated and the distribution of the renormalized transverse fields will approach the known form\cite{im}:
\begin{equation}
g(h) = \frac{d}{z}h^{-1+d/z}\;,
\label{h_distr}
\end{equation}
with some effective dynamical exponent, $z \approx z^{SR}(\theta_c)$ of the SR model at the control parameter $\theta_c$. The initial period of the RG ends when the generated new couplings become to be in the same order as the existing LR bonds. In the following \textit{intermediate period} dominantly transverse fields are decimated, but among the renormalized couplings - due to the maximum rule - there are more and more original LR bonds. As a result the distribution of the transverse fields will continuously change, the value of the effective dynamical exponent increases further and approaches its asymptotic value at the critical point, $z_c$. In the final, \textit{asymptotic regime} in the decimation mainly transverse fields are involved, but also a fraction, $r$,  LR couplings are decimated, too. (As the critical fixed point is approached $r$ tends to zero, the calculated scale-dependence is shown in Sec. \ref{numSDRG}.) The generated new couplings are almost always smaller than the existing LR bonds, thus according to the maximum rule these original couplings play the role of the renormalized ones. Consequently at the fixed point the decimation of a transverse field results in the erasing of the given site together with the couplings starting from it. The basic ingredients of the RG procedure in the three regimes are summarised in Table \ref{table_1}. The final cluster structure of the LR model in 1D is illustrated in Fig.\ref{fig_2}: comparing to the SR model here the clusters have smaller extent and contain less sites. In higher dimensions, in 2D and 3D we show in Fig.\ref{fig_3}
the structure of the largest clusters, in which the critical properties of the model are encoded. Even in higher dimensions these clusters are sparse and they can be embedded in a quasi-one-dimensional object. This last property is shared with the largest critical clusters in SR models.

\begin{table}[]
\centering
\caption{Properties of the critical SDRG procedure in the different regimes, see text.}
\label{table_1}
\begin{tabular}{|c|c|c|c|}
\hline
RG period    & decimation         & couplings          & $z_{\mathrm{eff}}$ \\ \hline
 initial     &  $h$               & SR                 & $\approx z^{SR}(\theta_c)$   \\
intermediate &  $h$               & SR and LR          & $z^{SR}(\theta_c) < z_{\mathrm{eff}} < z_c$  \\
 asymptotic  & $h$ and $J$        & LR                 & $z_c$ \\ \hline
\end{tabular}
\end{table}

\section{Numerical SDRG analysis}
\label{numSDRG}

\begin{figure}[th]
\begin{center}
\includegraphics[width=3.2in,angle=0]{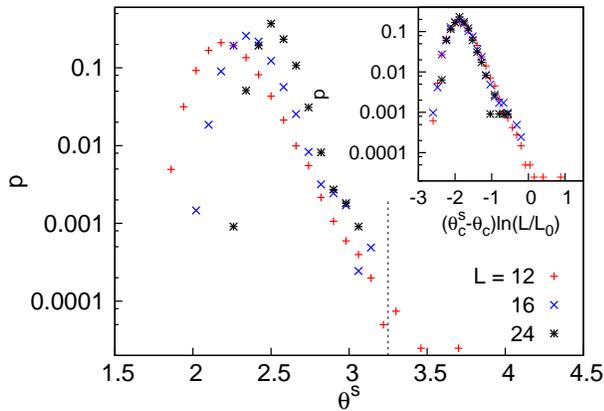}
\end{center}
\vskip -.5cm
\caption{
\label{fig_4} (Color online) Distribution of the pseudo-critical points, which are estimated to cross each other for different $L$ at $\theta_c \approx 3.25$, indicated by a dotted line. The ratio of the accumulated distributions on two sides of $\theta_c$ is given by $r_{\theta_c}(L)$, see the text and Fig.\ref{fig_4b}. The inset shows
the rescaled distributions with $L_0=2$.
}
\end{figure}

\begin{figure}[th]
\begin{center}
\includegraphics[width=3.2in,angle=0]{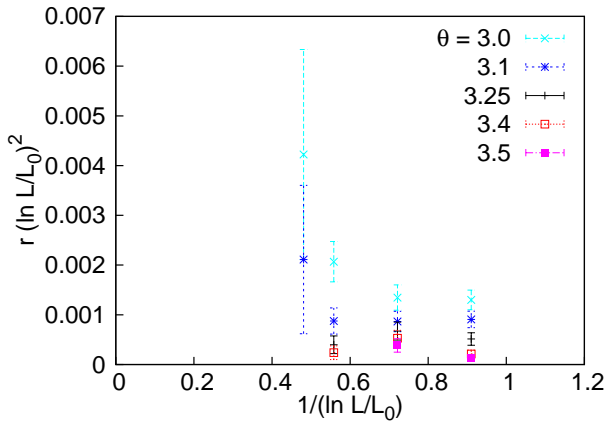}
\end{center}
\vskip -.5cm
\caption{
\label{fig_4b} (Color online) The scaled decimation ratio, $r_{\theta_c}(L)$, as a function of the $L$ size at $L_0=2$, indicating a similar logarithmic scaling as in 1D. 
}
\end{figure}

Here we present our numerical results for the three-dimensional LR model, which are obtained by the use of the fast SDRG algorithm in Ref.\cite{ddRG}. In the calculations we used finite samples with periodic boundary conditions of linear size up to $L=24$. The number of samples were
typically $100 000$ (at least $2000$ for the largest size) and the box-distributions are used, as described before.
We have fixed the decay exponent to $\alpha=4$ and calculated sample dependent pseudo-critical points, as described in Ref.\cite{2dki}. The distribution of the pseudo-critical points is shown in Fig.\ref{fig_4}: both the position of the maximum and the width of the distribution follows a $1/\ln L$ scaling, from which the true critical point is estimated at $\theta_c=3.25(15)$. The scaling behaviour of the distribution of the pseudo-critical points is compatible with an exponential increase of the correlation length, at least from the paramagnetic side:
\be
\xi \sim \exp\left(\rm const/(\theta-\theta_c)\right),\quad \theta > \theta_c\;.
\label{xi}
\ee
At the critical point we calculated the fraction of decimation steps, which involves a coupling and that, which involves a transverse field. They ratio, $r_{\theta_c}(L)$, is given by the ratio of the accumulated distributions of the pseudo-critical points on two sides of $\theta_c$. We obtained a logarithmic $L$-dependence: $r_{\theta_c}(L) \sim 1/(\ln L)^{\omega}$, with $\omega \approx 2$, as in the 1D case as illustrated in Fig. \ref{fig_4b}.

At the critical point, we have also calculated the average mass of the last remaining
cluster $\mu(L)=L^d m(L)$, $m(L)$ being the local magnetization and the characteristic time scale $\tau(L)$ defined as
$\tau=1/\tilde{h}$, where $\tilde{h}$ is the last decimated parameter in a finite sample.

\begin{figure}[th]
\begin{center}
\includegraphics[width=8cm]{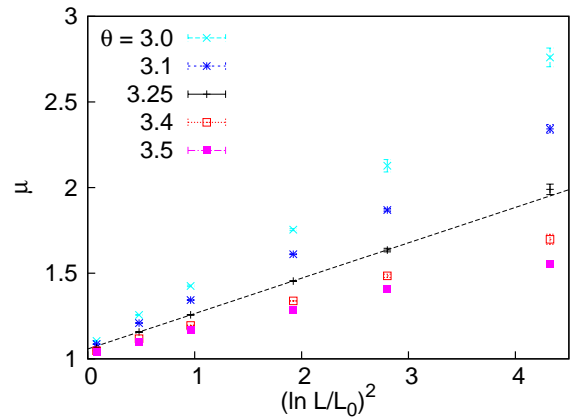}  
\end{center}
\vskip -.5cm
\caption{\label{fig_5} (Color online) The average mass of the last
  decimated cluster plotted against $(\ln L/L_0)^2$ with $L_0=3.2$.
The data has been
  obtained by numerical renormalization of the three dimensional model
  with decay exponent $\alpha=4$, for different values of the control
  parameter $\theta$.  }
\end{figure}

The numerical results indicate that the magnetic moment $\mu(L)$ has a slower-than-algebraic 
dependence, which can be written in analogy with the one-dimensional result as 
\be
\mu(L) \sim [\ln( L/L_0)]^{\chi}.
\label{mu_L}
\ee
Precise determination of $\chi$ from the existing
numerical results is difficult, since it is sensitive
to the value of the reference length, $L_0$.
The data in Fig.\ref{fig_5} are compatible with $\chi=2$ with $L_0=3.2$, but a similar fit is obtained
with $\chi=3$ if we choose $L_0=2$ instead.

\begin{figure}[th]
\begin{center}
\includegraphics[width=8cm]{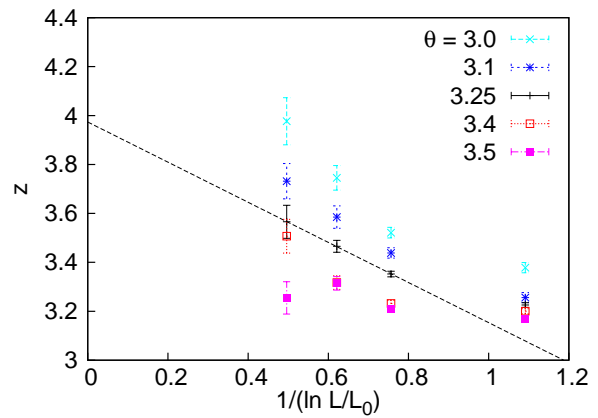} 
\end{center}
\vskip -.5cm
\caption{\label{fig_6} (Color online) 
Effective dynamical exponents obtained by two-point fits using Eq.(\ref{z_L}) as a function of the system size $L$. The straight line is a fit to the data obtained for the critical value, $\theta=3.25$. 
}
\end{figure}

Calculating the average logarithmic time scale, 
$\overline{\ln\tilde{h}}$,
estimates of an effective, size-dependent dynamical exponent,
$z(L)$, has been obtained from two-point fits of the relation
\be
\overline{\ln\tilde{h}}=-z\ln L+\mathrm{const}.
\label{z_L}
\ee
The extrapolation of $z(L)$ to infinite system size, as 
shown in Fig. \ref{fig_6}, is
compatible with the expectation $z_c=\alpha$. 

\begin{figure}[th]
\begin{center}
\includegraphics[width=8cm]{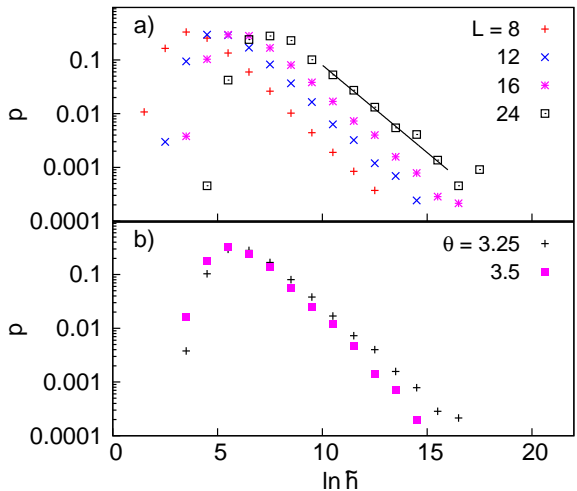} 
\end{center}
\vskip -.5cm
\caption{\label{fig_7} (Color online) 
a) Distributions of the last decimated transverse fields at the critical point for different sizes. The straight line indicates the asymptotic shape of the tail according to EVS, with $z=\alpha$, see text. b) As the $z$ dynamical exponent decreases, the shape of the distribution changes in the paramagnetic phase as illustrated at $L=16$.
}
\end{figure}

The dynamical exponent - according to Eq.(\ref{h_distr}) - is involved in the distribution of the last decimated transverse fields (see Eq.(\ref{frechet})), which is illustrated in Fig.\ref{fig_7}. At the critical point, see in Fig.\ref{fig_7}a, the numerical value of the critical dynamical exponent is compatible with $z_c=\alpha$. In the paramagnetic phase the distribution of the last decimated transverse fields is still in agreement with Eq.(\ref{h_distr}), but the dynamical exponent is $z<z_c$, see in Fig.\ref{fig_7}b.

\begin{figure}[th]
\begin{center}
\includegraphics[width=6.5cm,angle=0]{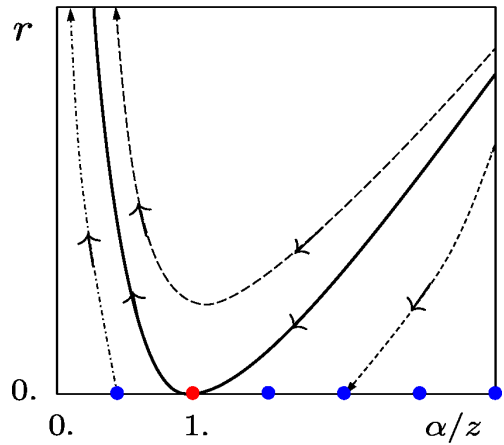}  
\end{center}
\vskip -.5cm
\caption{\label{fig_8} (Color online) Schematic SDRG phase diagram obtained through the maximum rule. At $r=0$ the attractive fixed points of the paramagnetic phase ($\alpha/z>1$) and the repulsive ones ($\alpha/z<1$) are separated by the critical fixed point, see the text.}
\end{figure}

We close this Section by presenting the SDRG phase diagram of the LR random transverse-field Ising model obtained with the maximum rule, see in Fig.\ref{fig_8}. Here we use the parameters, $\alpha/z$, 
and $r$, the ratio of the decimation frequencies of the couplings and the transverse fields. According to our numerical calculations the phase diagram has the same qualitative features in one-\cite{jki} and two-dimensions\cite{jki1}, too. (In one-dimension the phase-diagram is related to that of random Josephson-junctions\cite{akpr}). As seen in Fig.\ref{fig_8} there is a line of fixed-points at $r=0$, at which almost exclusively transverse fields are decimated. For $\alpha/z>1$ these fixed points are stable and control the paramagnetic phase and the corresponding Griffiths singularities\cite{griffiths}, while for $\alpha/z<1$ the fixed points are unstable and the RG-flow scales to $r \to \infty$, which corresponds to the ferromagnetic phase. In this regime the maximum rule in the SDRG procedure is certainly not valid. The two regimes of fixed points are separated by the critical fixed point at $\alpha/z=1$. In the following Section we analyse the scaling behaviour of the system in the vicinity of $r=0$ and $\alpha/z=1$ through extreme value statistics.

\section{Analysis of the critical behaviour trough EVS}
\label{EVS}

\subsection{Critical point}

At the critical point in the asymptotic regime of the RG mainly transverse fields are decimated, but occasionally LR bonds are decimated, too. Let us consider a finite system of linear length, $L$, and concentrate on the largest cluster, which contains $\mu+1$ sites. In this cluster altogether $\mu$ LR bonds \textit{have been decimated} and let us denote them by: $J_i=b_i r_i^{-\alpha}$ where $i=1,2,\dots,\mu$ is the order of decimation. In this cluster the transverse-fields $h_i$, $i=1,2, \dots,\mu+1$ \textit{are not decimated} out, thus $J_i \gtrsim h_i$. $J_i$ being the only decimated coupling in a region of linear size $r_i$, while $h_i$ is the smallest one out of $\sim r_i^d$ transverse fields. Since the transverse fields in the asymptotic regime have a power-law distribution, see in Eq.(\ref{h_distr}) the $h_i$ is given by EVS as $h_i \simeq \kappa_i r_i^{-z}$ where $\kappa_i$ are random numbers which are distributed according to the Fr\'echet statistics:
\begin{equation}
P(\kappa)=\frac{d}{z}\kappa^{d/z-1} \exp(-\kappa^{d/z})\;.
\label{frechet}
\end{equation}
The effective transverse field of the cluster is given by: $\tilde{h} \sim \prod_{i=1}^{\mu+1} h_i/\prod_{i=1}^{\mu} J_i \sim h_{\mu+1}\prod_{i=1}^{\mu} \left( r_i^{\alpha-z}\kappa_i/b_i \right)$. This scales differently for $\overline{\ln(J)} > \overline{\ln(h)}$ and for $\overline{\ln(J)} < \overline{\ln(h)}$, where the overbar denotes averaging over disorder. Thus at the critical point $\alpha=z_c$ and
$\overline{\ln(b)} = \overline{\ln(\kappa)}$. This result about the dynamical exponent at the critical point agrees with our numerical results in the previous section. For the given cluster at the critical point the effective transverse field is given by: $\tilde{h} \sim\prod_{i=1}^{\mu}\left( \kappa_i/b_i \right)$ (since $h_{\mu+1}={\cal O}(1)$). If $\kappa_i$ and $b_i$ are not (or just weakly) correlated, then according to the central limit theorem $\ln \tilde{h} \sim \mu^{1/2}$. More generally we can write $\ln \tilde{h} \sim \mu^{1/\chi}$, what
is to be compared with $\ln \tilde{h} \sim -\alpha \ln L$, which implies $\mu \sim (\ln L)^{\chi}$, in agreement with Eq.(\ref{mu_L}).

\subsection{Paramagnetic side}

In the paramagnetic phase, $\theta > \theta_c$, the distribution of the transverse fields in Eq.(\ref{h_distr}) involves the dynamical exponent $z < \alpha$ and in the vicinity of the transition point $\alpha-z=\delta \alpha/d$ and $\delta \sim \theta/\theta_c-1 \ll 1$. Here the correlation length, $\xi(\delta)$, is defined by the length of the longest decimated bond, $r_l$, the corresponding coupling, $J_l=b_l r_l^{-\alpha}$ being larger than the smallest transverse field: $h_l \simeq \kappa_l r_l^{-\alpha(1-\delta/d)}$. Consequently $\kappa_l < \xi^{-\delta \alpha/d}$
or $\mathrm{Prob}(\kappa_l < \xi^{-\delta \alpha/d})={\cal O}(1)$, which can be written using Eq.(\ref{frechet}) as:
\begin{equation}
\int_0^{\xi^{-\delta \alpha/d}}P(\kappa)\mathrm{d} \kappa=1-\exp(-\xi^{-\delta}) \approx \xi^{-\delta}={\cal O}(1)\;,
\end{equation}
for $z \approx \alpha$.
Consequently in the paramagnetic phase the correlation length is given by: $\ln \xi \sim 1/\delta$, in agreement with Eq.(\ref{xi}).

The dynamical behaviour of the system in the paramagnetic phase is governed by Griffiths singularities, which are due to rare regions, in which the system is locally in its ferromagnetic phase. Here we recapitulate the so called optimal fluctuation argument\cite{thill_huse} for the SR model and then generalise it to the LR case. The probability, $P(\ell)$ to find an ordered domain of linear size $\ell$ in the paramagnetic phase is exponentially small: $\ln P(\ell) \sim (\ell/\xi)^d$, but its excitation energy, $\epsilon_{SR}(\ell)$ - estimated through an $\ell^d$-order perturbation theory in $h_i/J_{ij}$- is also exponentially small: $\ln \epsilon_{SR}(\ell) \sim (\ell/l_0)^d$. Combining these two effects a power-law distribution of $\epsilon_{SR}$ is observed, with a dynamical exponent $z_{SR}=d(\xi/l_0)^d$. In the LR model we first assume that the rare regions are localised, too, then the form of $P(\ell)$ remains the same, however by estimating the excitation energy one should take into account the LR forces, too. Let us assume that the ordered cluster of linear size $\ell$ is the largest one (thus has the smallest excitation energy) within a region of linear size $L$, where $L^d P(\ell)={\cal O}(1)$, thus $\ln L \approx \frac{1}{d}(\ell/\xi)^d$. Within this controlled region the distance between the largest and the second largest clusters is $\sim L$ and the direct LR interaction between them is $\epsilon_{LR} \sim L^{-\alpha}$. Now in the LR model the actual value of the excitation energy in the controlled domain is obtained by comparing the SR and LR contributions and is given by $\epsilon(\ell)=\mathrm{max}[\epsilon_{SR}(\ell),\epsilon_{LR}(L)]$, or $\ln \epsilon(\ell)=\mathrm{max}[-(\ell/l_0)^d,-\alpha/d (\ell/\xi)^d]$. This means, that the effective volume-scale in the SR model, $l_0^d$, is replaced by $\mathrm{max}[\frac{d}{\alpha}\xi^d,l_0^d]$ in the LR model. Consequently the dynamical exponent in the LR model is given by: $z=\mathrm{min}[z_{SR},\alpha]$, i.e. it is bounded by $z=z_c=\alpha$, which is the value at the critical point, as observed numerically. Our numerical results in Sec.\ref{numSDRG} are in favour of our assumption that the rare regions in the LR models are localised, too.

\subsection{Ferromagnetic side}

In the ferromagnetic phase, $\theta < \theta_c$, we analyse the SDRG solution with the maximum rule, starting from the unstable fixed points in the vicinity of the critical fixed-point, see in Fig. \ref{fig_8}. We expect that the asymptotic results in the vicinity of the transition point do not depend on the actual direction, how the transition point is approached from the ferromagnetic phase. In these fixed points the distribution of the transverse fields is given in Eq.(\ref{h_distr}) with $z=\alpha(1+\delta/d)$ and $\delta \sim 1-\theta/\theta_c \ll 1$. In the ferromagnetic phase there is a giant connected cluster and the length-scale, $\xi$, is defined by the linear extent of the largest hole in it. This is defined by the length of the longest decimated bond, $r_l$,
so that all the transverse fields are decimated out within this region. The strength of this bond now satisfies: $J_l/h_l \sim r_l^{\delta \alpha/d}/\kappa_l<1$. This means, that $\mathrm{Prob}(\kappa_l > \xi^{\delta \alpha/d})={\cal P}={\cal O}(1)$ and here we assume once more, that $P(\kappa)$ can be described by the Fr\'echet distribution:
\begin{equation}
\int_{\xi^{\delta \alpha/d}}^{\infty}P(\kappa)\mathrm{d} \kappa=\exp(-\xi^{\delta}) = \exp(-e^{-C})={\cal P}\;,
\end{equation}
thus $\xi \sim \exp(-C/\delta)$. Evidently $\xi$ has different scaling behaviour for $C<0$ ($0 < {\cal P} < 1/e$) and for $C>0$ ($1/e < {\cal P} < 1$). In the former case $\xi$ is divergent for $\delta \to 0$ as in the paramagnetic side, but for $C>0$ the correlation length in the continuum description goes to zero and the magnetization is of ${\cal O}(1)$ for $\delta \to 0$. Here - depending on the possible set of values of ${\cal P}$ in the different samples - we can have two different scenarios concerning the behaviour of the average magnetization at the transition point.

\subsubsection{Second-order transition}

If, due to some reason, ${\cal P}$ is smaller than $1/e$ in every sample, then we always have $C<0$ and the average magnetization vanishes at $\delta \to 0$, thus the phase transition is of second order. In this case the average correlation length diverges exponentially, as in the paramagnetic side.

\subsubsection{Mixed-order transition}
\label{mixed-order}

If, however, ${\cal P}$ is not bounded by $1/e$ and there is a finite fraction of the samples with $1/e < {\cal P} < 1$, thus $C>0$, then the average magnetization goes to a finite limiting value as the transition point is approached from the ferromagnetic side. At the same time the average correlation length is exponentially divergent, thus the transition is of mixed order. 

At the moment we have no information about the possible values of ${\cal P}$, and in the lack of any known constrains we incline to prefer the mixed-order transition scenario. It is in the spirit of Occam's razor, since in this case we have to use fewer assumptions. Mixed-order transitions often appear in pure systems with LR forces\cite{anderson1969exact,thouless1969long,dyson1971ising,cardy1981one,aizenman1988discontinuity,slurink1983roughening,bar2014mixed} our model then would represent such a phenomena in the presence of random LR interactions.

\section{Discussion}
\label{discussion}
In this paper we have studied the critical properties of the random transverse-field Ising model in 3D with the presence of LR forces. Our present work completes our investigations starting in 1D\cite{jki} and having announced some numerical results in 2D\cite{jki1}. This problem is technically quite difficult and the only possible method of numerical investigations at present days seems to be the SDRG approach. Here we used a variant of it based on the so called maximum rule, which enabled us to study sufficiently large systems (up to linear size L=24) with an appropriate statistics. The obtained RG phase-diagram in Fig. \ref{fig_8} has the same qualitative structure as that in lower dimensions, both the phase-transition point and the fixed points controlling the paramagnetic phase with Griffiths singularities are located at $r=0$ and are characterised by parameter dependent dynamical exponents. In these attractive fixed points almost exclusively transverse fields are decimated and the renormalised couplings, according to the maximum rule are selected from the original LR bonds and their value can be estimated from EVS. We have found that the average correlation length at the transition point diverges exponentially, see Eq.(\ref{xi}), but the average magnetisation - in the spirit of Sec.\ref{mixed-order} - is expected to have a finite jump at the transition point. Therefore we conjecture that the transition is of mixed-order. Mixed-order transitions have already been observed in different systems: in the classical LR Ising-chain with $\alpha=2$\cite{anderson1969exact,thouless1969long,dyson1971ising,cardy1981one,aizenman1988discontinuity,slurink1983roughening,bar2014mixed}; in models of depinning transition\cite{PS1966,fisher1966effect,blossey1995diverging,fisher1984walks}.; and in percolation models with glass and jamming transition\cite{gross1985mean,toninelli2006jamming,toninelli2007toninelli,schwarz2006onset,liu2012core,liu2012extraordinary,zia2012extraordinary,tian2012nature,bizhani2012discontinuous,sheinman2014discontinuous}, for a recent review see in Ref.\cite{bar2014mixed1}. Our study indicates that such a phenomenon could take place also in disordered quantum systems with LR forces.

At the transition point the magnetisation in a finite sample scales logarithmically, see Eq.(\ref{mu_L}) and the dynamical exponent is finite, $z_c=\alpha$, thus the critical fixed point is a strong disorder fixed point, but not an infinite disorder one. Therefore the obtained SDRG results are not asymptotically exact, however these are very probably qualitatively correct and also the numerical estimates are expected to be reliable, as already noticed in other systems having a strong disorder fixed point\cite{lin}. Considering the application of the maximum rule in our numerical algorithm it has negligible effect in the paramagnetic phase, in which the typical size of ferromagnetic clusters is finite. At the transition point we expect to have at most logarithmic corrections to the results obtained by the maximum rule. Finally, at the ferromagnetic phase the maximum rule does not hold any longer, but the predicted jump of the magnetisation at the transition point most probably remains true together with the mixed-order nature of the transition.

Our results are expected to hold for a large class of disordered LR quantum models having a discrete symmetry, such as the random quantum Potts and Ashkin-Teller models\cite{im}. The critical fixed-point of the model in Eq.(\ref{eq:H}) is expected to govern the critical behaviour of some random stochastic models, such as the random contact process, at least for strong enough disorder. For SR forces this type of mapping is known quite some time\cite{hiv,vd}, and its validity has also been demonstrated with LR interactions in 1D and 2D\cite{jki1,note}.

\begin{acknowledgments}
This work was supported by the National Research Fund under grant
No. K109577.
\end{acknowledgments}


\begin{thebibliography}{99}
\bibitem{note} For the random LR contact process the discontinuity of the order-parameter at the transition point is difficult to demonstrate numerically, at least for not too strong disorder. Here probably there is a disorder dependent cross-over, as has been seen in the SR model\cite{hiv}.
\end{thebibliography}
\end{document}